\documentclass[%
pra%
,reprint%
,twocolumn%
,secnumarabic%
,floatfix%
,superscriptaddress, showpacs%
,amssymb, amsmath, nobibnotes, aps]{revtex4-1}

\usepackage[latin1]{inputenc}
\usepackage{epsfig}
\usepackage{color}

\newcommand{\ind}[1]{_{\text{#1}}}
\newcommand{\bra}[1]{\left<#1\right|}
\newcommand{\ket}[1]{\left|#1\right>}

\begin{document}

\title{Dynamic formation of Rydberg aggregates at off-resonant excitation}

\author{Martin G\"{a}rttner}
\affiliation{Max-Planck-Institut f\"{u}r Kernphysik, Saupfercheckweg 1, 69117 Heidelberg, Germany}
\affiliation{Institut f\"{u}r Theoretische Physik, Ruprecht-Karls-Universit\"{a}t Heidelberg, Philosophenweg 16, 69120 Heidelberg, Germany}
\affiliation{ExtreMe Matter Institute EMMI,
              GSI Helmholtzzentrum f\"ur Schwerionenforschung GmbH,
              Planckstra\ss e~1,
              64291~Darmstadt, Germany}

\author{Kilian P.~Heeg}
\affiliation{Max-Planck-Institut f\"{u}r Kernphysik, Saupfercheckweg 1, 69117 Heidelberg, Germany}

\author{Thomas Gasenzer}
\affiliation{Institut f\"{u}r Theoretische Physik, Ruprecht-Karls-Universit\"{a}t Heidelberg, Philosophenweg 16, 69120 Heidelberg, Germany}
\affiliation{ExtreMe Matter Institute EMMI,
              GSI Helmholtzzentrum f\"ur Schwerionenforschung GmbH,
              Planckstra\ss e~1,
              64291~Darmstadt, Germany}

\author{J\"{o}rg Evers}
\affiliation{Max-Planck-Institut f\"{u}r Kernphysik, Saupfercheckweg 1, 69117 Heidelberg, Germany}

\date{\today}

\pacs{67.85.-d,32.80.Ee,42.50.Nn}

\begin{abstract}
The dynamics of a cloud of ultra-cold two-level atoms is studied at off-resonant laser driving to a Rydberg state. We find that resonant excitation channels lead to strongly peaked spatial correlations associated with the buildup of asymmetric excitation structures. These aggregates can extend over the entire ensemble volume, but are in general not localized relative to the system boundaries. The characteristic distances between neighboring excitations depend on the laser detuning and on the interaction potential. These properties lead to characteristic features in the spatial excitation density, the Mandel $Q$ parameter, and the total number of excitations. As an application an implementation of the three-atom CSWAP or Fredkin gate with Rydberg atoms is discussed. The gate not only exploits the Rydberg blockade, but also utilizes the special features of an asymmetric geometric arrangement of the three atoms.
We show that continuous-wave off-resonant laser driving is sufficient to create the required spatial arrangement of atoms out of a homogeneous cloud. 
\end{abstract}

\maketitle

\section{Introduction}

During the last few years an immense research activity both experimentally and theoretically  aimed at a detailed understanding of the properties of ultra-cold gases whose atoms are excited to states of high principal quantum number $n$. Such highly excited Rydberg atoms have extreme properties which make them promising candidates for a number of fascinating applications~\cite{saffman2010,comparat2010}. Most importantly, they feature  long-range van der Waals (vdW) interactions, which lead to interesting effects like the dipole blockade~\cite{lukin2001, jaksch2000}. 
After initial work mainly concerned with bulk properties of Rydberg ensembles, current effort focuses more and more on spatially resolved observations. This is motivated not least by experimental progress~\cite{schwarzkopf2011,schauss2012} and theoretical proposals~\cite{olmos2011,guenter2012} for spatially resolved excitation imaging. Also on the theoretical side, in particular rate equation models~\cite{ates2007a, heeg2012, petrosyan2012b, hoenig2013} and full many-body simulations on truncated Hilbert spaces~\cite{robicheaux2005, ates2006, younge2009, weimer2008, weimer2010b, loew2009, olmos2009a, tezak2011, mayle2011, breyel2012, lee2011} provide a handle to access spatially resolved properties. 

In this spirit,  it has been predicted that spatial pair correlations can be induced in a three-level Rydberg gas via the so-called anti-blockade arising from resonant excitations due to single-atom Autler-Townes splitting~\cite{ates2007b}. These correlations could be measured based on mechanical forces due to the vdW interaction. The induced particle motion leads to an encoding of position correlations into a time-dependent Penning ionization signal~\cite{amthor2010}. Interestingly, in this way, spatial information is gained without a spatially resolved measurement.
Following these works on spatial properties of pairs of Rydberg atoms in large ensembles, it was subsequently shown that also the Rydberg ensemble as a whole could form crystalline structures in the Rydberg excitation density~\cite{pohl2010, vanbijnen2011, sela2011}. In a repulsive vdW gas of Rydberg atoms,  it can be energetically favorable for a given number of excitations to assume a highly ordered, crystalline state, with distances between the excitations maximized. If the laser detuning acts opposite to the interaction contribution, these ordered states  become the quantum mechanical ground state of the system. It was proposed that such ground-state crystals (GSC) could be produced using chirped laser pulses~\cite{pohl2010, vanbijnen2011, schachenmayer2010}. The chirp of the laser driving induces adiabatic passage to the energetically most favorable state corresponding to the crystalline excitation structure.
For dipole-dipole interacting spins in a lattice configuration a scheme for growing ordered structures by using resonant excitation processes similar to the ones discussed here has been proposed recently \cite{lemeshko2012}.

Here, we study regular excitation structures that arise in an off-resonantly driven disordered gas at fixed laser detuning. We analyze in detail the spatially resolved properties of a one-dimensional disordered gas of Rydberg atoms via numerical simulations on a truncated Hilbert space. The regular structures or aggregates predicted here are fundamentally different from the GSC reported previously. First, the characteristic distances between excitations do not depend on the trap geometry, but only on the laser detuning and interaction potential. Second, the structures are not spatially localized with respect to the trap, but can float over a certain position range between different realizations. Third, global observables, such as the excitation number or the Mandel $Q$ parameter yield results qualitatively different from those for the GSC. Also, due to the floating nature of the aggregates, the total and the excitation-number resolved spatial excitation densities exhibit characteristic structures. Methods to 
experimentally 
verify the 
aggregate structure  are discussed. The underlying resonant excitation mechanism is dynamical, inviting time-resolved studies of the formation of spatial correlations. Part of our predictions can already be probed through the detuning and density dependence of the bulk total number of excitations without the need for spatial resolution. 

We find that two characteristic lengths arise in an off-resonantly excited gas. These length have a fixed ratio of $2^{1/d}$ for an interaction potential $V\sim 1/r^d$ and can be explained by two different resonant excitation channels connected to one-photon and two-photon processes, respectively. As an application, we show this unique property can be used to construct a quantum gate. We find that the gate fidelity sensitively depends on the specific geometric arrangement of the excited atoms. High fidelity is achieved, as our resonant excitation scheme automatically self-assembles the optimum arrangement out of a homogeneous cloud of atoms, independent of the laser parameters.

\section{Model and Hamiltonian}

Our model system is a one-dimensional cloud of Rydberg atoms in two-level and frozen-gas approximations~\cite{comparat2010}. We assume temporally and spatially constant laser intensity and wavelength. The corresponding many-body Hamiltonian in a suitable interaction picture and rotating wave approximation reads ($\hbar=1$) \cite{robicheaux2005,comparat2010,younge2009}
\begin{equation}
\label{eq:Hamiltonian}
 H=\sum_{i=1}^N \left[ -\Delta s^{(i)}_{ee} + \frac{\Omega}{2}\sigma_x^{(i)}\right]
    + C_6 \sum_{i<j}^N \frac{s^{(i)}_{ee}s^{(j)}_{ee}}{r_{ij}^6} \,,
\end{equation}
with $s^{(i)}_{\alpha\beta}=\ket{\alpha}_i\bra{\beta}$ and $\sigma_x^{(i)}=s^{(i)}_{eg} + s^{(i)}_{ge}$ for atom $i$. The first part of this Hamiltonian contains a sum over single-atom contributions. It includes the detuning $\Delta$ between laser frequency and atomic transition frequency, and the laser coupling between the ground state $\ket{g}$ and Rydberg state $\ket{e}$ with Rabi frequency $\Omega$. The second part accounts for the vdW interactions between two atoms in the Rydberg state at a mutual distance $r_{ij}$. 
We numerically solve the time dependent Schr\"{o}dinger equation with Hamiltonian (\ref{eq:Hamiltonian}) for given atom positions $\mathbf{r}_i$ and parameters $C_6$ , $\Omega$ and $\Delta$, starting from an initial state with all atoms in the ground state. 
%
For a many-body system, a naive integration is impractical, as the number of states grows exponentially with the particle number. We overcome this problem by exploiting the Rydberg blockade to truncate the Hilbert space to a physically relevant subspace~\cite{gaerttner2012,younge2009}. 
All simulation results are checked for convergence with respect to the truncation parameters. Additionally, to obtain a convergence of the ensemble state with time, a Monte Carlo sampling over runs with different atom positions is employed~\cite{gaerttner2012,younge2009, carroll2009, ryabtsev2010}. In all the simulations shown here, the Rabi frequency was kept at $\Omega = 10$\,MHz, and the interaction strength at $C_6 = 6.6$\,THz\,$\mu$m$^6$. This choice, however is not restrictive since the results can be mapped to any other parameter values by rescaling time and length. For these parameters, the systems studied in this work would typically converge at physical evolution times of about 2\,$\mu$s, while the reported observables have been obtained at time 5\,$\mu$s.

\section{Results}

\subsection{Spatial correlations}

\begin{figure}[t]
 \centering
 \includegraphics[width=\columnwidth]{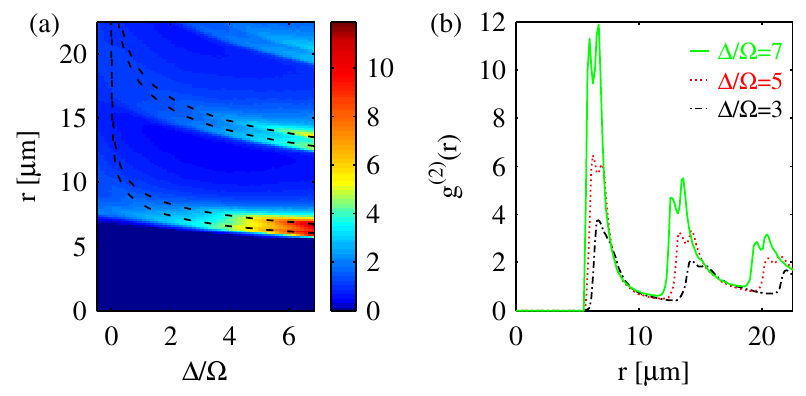}
 \caption{(Color online) (a) Pair correlation function $g^{(2)}(r)$ as a function of interparticle distance $r$ and detuning $\Delta$. Dashed lines are theoretical prediction for resonance positions as explained in the main text. (b) Sections through (a) at detunings $\Delta/\Omega=3,5,7$. Parameters are $L=30\,\mu$m, $N=30$, $\Omega=10\,$MHz, and $C_6=6.6\,$THz\,$\mu$m$^6$.}
 \label{fig:g2}
\end{figure}

\begin{figure}[t]
 \centering
 \includegraphics[width=\columnwidth]{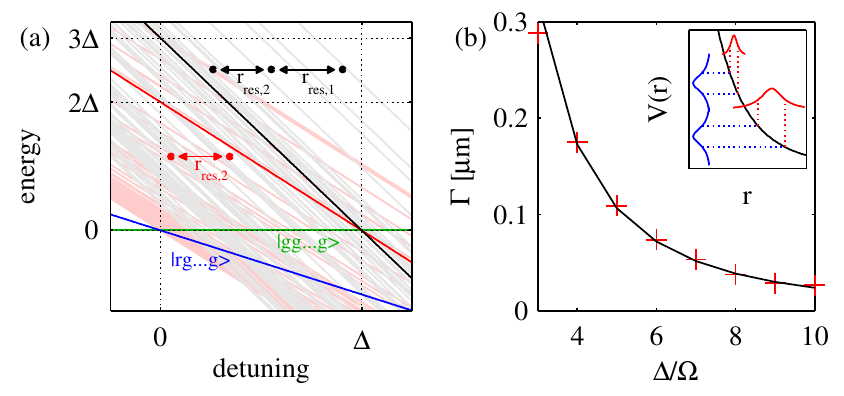}
 \caption{(Color online) (a) Illustration of the resonant excitation channels. Bare state energies [diagonal part of the Hamiltonian \eqref{eq:Hamiltonian}] as a function of the laser detuning. We used a sample of $16$ atoms randomly placed in a one-dimensional trap. The energy of $m$-fold excited states decreases with slope $m$. The solid red (upper gray) and black lines mark states that can be resonantly coupled to from the ground state for a certain detuning $\Delta$. Note that in a dressed state picture (eigenstates of the Hamiltonian) all the crossings would become avoided ones. (b) Dependence of the two-photon excitation linewidth on the detuning. The solid line is obtained from Eq.~(\ref{eq:res_width}), red crosses are obtained from the numerical simulations. The inset shows the detuning-dependent transformation of energetic resonances into spatial resonances via the interaction potential $V(r)=C_6/r^6$.}
 \label{fig:bare_states_pot}
\end{figure}

Here we show that characteristic geometric excitation structures form out of the homogeneous cloud of atoms under continuous laser driving. For this we study the pair correlation function
\begin{equation}
 g^{(2)}(r) = \frac{1}{N_p(r)}{\sum_{i,j}}' 
	      \frac{\langle s^{i}_{ee} s^{j}_{ee} \rangle}{\langle s^{i}_{ee} \rangle \langle s^{j}_{ee} \rangle}\, ,
\end{equation}
where $\sum_{i,j}'$ sums over pairs with distance in $[r,r+\Delta r]$  and $N_p(r)$ is the total number of such pairs. $g^{(2)}(r)$ is a measure for the conditioned probability for an excitation if another excitation is already present at a distance $r$. As shown in Fig.~\ref{fig:g2}, pronounced resonances emerge in $g^{(2)}$ at regularly spaced distances for detunings $\Delta > \Omega$.
At large detunings ($\Delta/\Omega\gg 1$) and without the interaction ($C_6 \to 0$), the laser field is off-resonant with all transitions and does not induce excitations. But if the vdW interaction is included, it can shift transitions to higher excited states into resonance with the laser~\cite{ates2007b,amthor2010,mayle2011,lemeshko2012}.  
A closer look at Fig.~\ref{fig:g2} reveals a sequence of double resonances. The first resonance at lowest interparticle distance is the well-known resonant pair excitation~\cite{ates2007b} in which the ground state is resonantly coupled to a doubly excited state via two-photon excitation if $2\Delta = C_6/r\ind{res,2}^6$. We identified the second maximum at slightly higher $r$ as the resonant transition from an $m$-fold to an $m$+1-fold excited state, satisfying $\Delta = C_6/r\ind{res,1}^6$. The two conditions are shown as dashed lines in Fig.~\ref{fig:g2}(a) and coincide very well with the maxima of $g^{(2)}$ from the numerical simulation results. We can thus directly trace back the emergence of spatial correlations to resonant excitation channels.
Starting from a doubly-excited state with interatomic distance $r\ind{res,2}$, a triply excited state emerges. This is also illustrated in Fig.~\ref{fig:bare_states_pot}(a) where we show the bare energies (diagonal elements of the Hamiltonian) as a function of the detuning. It can be seen that, as long as the density of states is sufficiently high around zero energy, pair states and triplet states are available that can be excited resonantly. For the later analysis it is crucial to note that due to the mutual interaction shifts, the third atom has distance $r\ind{res,1}\neq r\ind{res,2}$, such that an asymmetric three-particle structure is created. This triplet causes the third resonance line at $r\ind{res,1}+ r\ind{res,2}$ in Fig.~\ref{fig:g2}(a). Subsequent resonances originate from higher-excited states, which most likely occur again at distances  $\approx r\ind{res,1}$ from the respective previous structures. In total, for highly excited states, a regular chain of atoms with a single ``defect'' formed by 
the initial pair of atoms is created.

Interestingly, the triply excited state is distinguished by a ratio $r\ind{res,1}/r\ind{res,2}=2^{1/d}$ for an interaction potential $V\sim 1/r^d$, independent of the trap size and the laser parameters. In this sense, the interaction potential leads to a self-assembly of asymmetric excitation structures. This invites applications exploiting the robust and definite asymmetric spatial configuration with distances $r\ind{res,1}$ and $r\ind{res,2}$ between the excitations, created out of a homogeneous cloud of atoms. An example for this will be discussed in Sec.~\ref{sec:gate}.

\begin{figure}[t]
 \centering
 \includegraphics[width=\columnwidth]{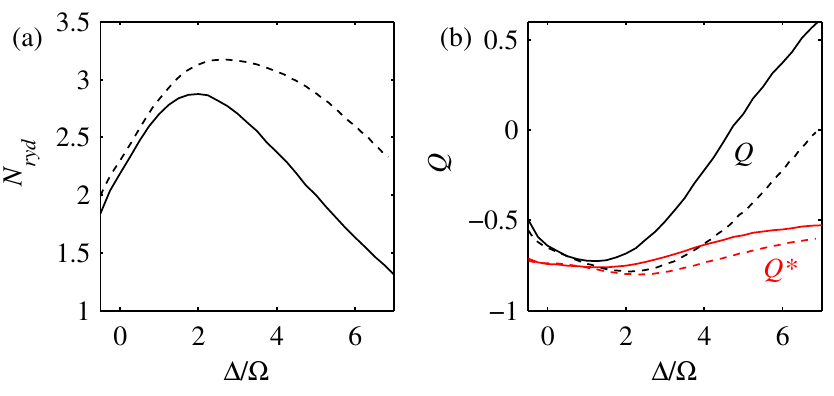}
 \caption{(Color online) (a) Total number of excited atoms and (b) Mandel $Q$ parameter as a function of the detuning. $Q^*$ is the corresponding parameter evaluated without the ground state fraction. Solid lines show $N=30$; dashed lines show $N=45$. $C_6$ and $L$ are as in Fig.~\ref{fig:g2}.}
 \label{fig:Q_Nex}
\end{figure}

Next, we analyze the structure of the spatial correlations at high detunings in more detail. Figure~\ref{fig:g2}(b) shows that the pair correlation resonances become narrower with increasing detuning. The origin of this effect is illustrated in the inset of Fig.~\ref{fig:bare_states_pot}(b). The resonances are of Lorentzian shape in energy space as a function of the detuning. The interaction potential $V(r)$ translates this dependence on the detuning into a distance dependence. For the same energy-resonance width, the position-space resonances become narrower as the distance decreases at which the laser moves into resonance. 
Hence, an increase of the detuning leads to a reduction of the resonant width. 
For the particular case of a two-photon resonance, the energetic resonance width depends on the detuning itself. To quantify this, we adiabatically eliminate the singly excited states in the two-atom problem. This is valid if $\Delta$ is much larger than both, $\Omega$ and the two photon detuning $V-2\Delta$. The resulting effective Rabi frequency is $\Omega^2/\Delta$, which means that the full width of the resonance with respect to the two photon detuning (power broadening) is $2\Omega^2/\Delta$.
Linearizing the potential around $r\ind{res,2}$ we obtain the position-space width of the two photon resonance 
\begin{equation}
\label{eq:res_width}
 \Gamma = \frac{2\Omega^2}{\Delta}\left(\frac{dV}{dr}\right)_{r=r\ind{res,2}}^{-1} = \frac{1}{6} \frac{\Omega^2}{\Delta^2} \left(\frac{C_6}{2\Delta}\right)^{1/6} \, .
\end{equation}
As shown in Fig.~\ref{fig:bare_states_pot}(b), the numerical simulation data fully agrees with Eq.~(\ref{eq:res_width}) up to an overall prefactor of about 2 due to additional atoms in the trap which broaden the resonance compared to the idealized two-atom case. Note that, since the potential is not strictly linear across the width of the resonance, the spatial shape of the resonance peaks is asymmetric.

In summary, we find that in the limit of large $\Delta$, the position-space two-photon resonance width decreases as $\Delta^{-13/6}$. Therefore, in our model, resonant excitations are only possible at definite positions with lattice spacings determined by the shape of the interaction potential. In the limit of large detuning all features of the pair correlation function can be understood in terms of two-atom properties.

\subsection{Excitation density and excitation statistics}

\subsubsection{Number of Rydberg excitations}

After having established the formation of stronger spatial ordering with increasing detuning, we now address the question how this ordering manifests itself in various observables related to the spatial distribution of Rydberg excitations. 
We start with the number of Rydberg excitations $N\ind{ryd}$. Fig.~\ref{fig:Q_Nex}(a) shows that $N\ind{ryd}$ increases with detuning starting from the resonant case, at some positive detuning assumes a maximum, and then decreases towards larger values of $\Delta$. The initial increase can be attributed to the presence of the off-resonant excitation channels at positive detuning, which also lead to a better packing of the Rydberg excitations. But with increasing detuning, the spatial excitation resonances become more narrow, as shown in Fig.~\ref{fig:bare_states_pot}(b). Therefore, the number of atom pairs with distance compatible with the resonant distance reduces. Consequently, starting from a certain critical detuning, the total number of excited atoms decreases with increasing $\Delta$.
For higher densities, this effect is expected to set in at higher detunings, since at a given resonance width,  the number of pairs having a distance within the resonant range increases with density. This is supported by our numerical simulation data, as the maximum  of the number of Rydberg excitations $N\ind{ryd}$ is shifted to higher detunings for higher densities, see Fig.~\ref{fig:Q_Nex}(a). 

\subsubsection{Rydberg excitation statistics}

Next we investigate the distribution of the number of excitations, which  is characterized by the Mandel $Q$ parameter
\begin{equation}
 Q = \frac{\langle \hat{N}\ind{ryd}^2\rangle - \langle \hat{N}\ind{ryd}\rangle^2}{\langle \hat{N}\ind{ryd}\rangle} - 1.
\end{equation}
For a Poissonian distribution of excitation numbers, like in a coherent state, the $Q$ parameter is zero, while for sub-Poissonian statistics it is negative. 
In previous proposals on GSC, for a given detuning, a definite number of excitations was predicted in the system. Then, ideally $Q$ reaches the value $-1$ for a Fock state.

In contrast, Fig.~\ref{fig:Q_Nex}(b) shows that in our setup, the $Q$ parameter assumes a minimum at a detuning close to that of the maximum in $N_{ryd}$, but then increases  again towards larger $\Delta$. The interpretation of this result is related to the behavior of $N_{ryd}$. Initially, the off-resonant  channels for ordered excitation structures lead to  a decrease of $Q$. 

But towards larger detunings, i.e., smaller resonance widths, there is an increasing probability that
there are few or even no atom pairs at the resonant distance in the ensemble. Furthermore, the spatial width of the resonant pair excitation (aggregate nucleation) decreases faster with the detuning than the width for subsequent resonant excitation of further atoms (aggregate growth). Therefore, once a resonant pair is  excited, there is a high probability that further atoms connected to this initial seed are excited subsequently. This then leads to a non-zero probability to detect no exciation at all and at the
same time to the emergence of aggregates of size two and higher, while the population of singly excited states is very low. A signature for this effect is that at large detunings, the system evolves into a bimodal excitation distribution, with one fraction in the ground state, and the second fraction distributed around a parameter-dependent excitation number. This observation is supported by corresponding results shown in Fig.~\ref{fig:Q_Nex}(b) for $Q^*$, which is the ordinary $Q$ parameter evaluated without the ground state fraction. It can be seen that $Q^*$ remains low even for higher detunings $\Delta$.

We thus conclude that in the present case of off-resonantly excited Rydberg gases, super-Poissonian rather than sub-Poissonian excitation number statistics are a sign of ordered structures and the buildup of strong correlations, in contrast to the properties of the GSC.

\begin{figure}[t]
 \centering
 \includegraphics[width=\columnwidth]{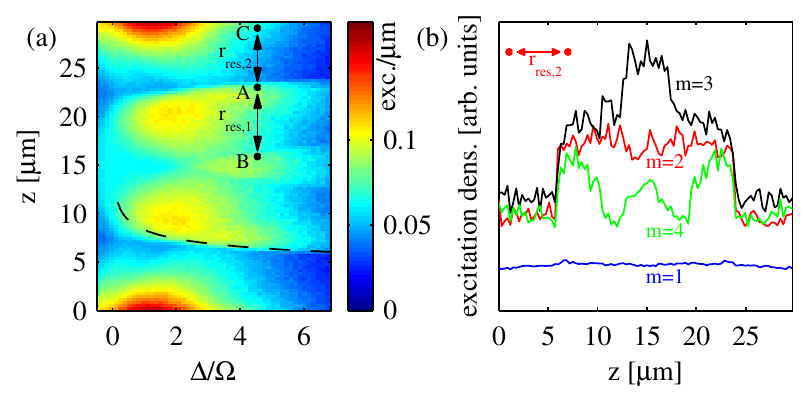}
 \caption{(Color online) (a) Rydberg excitation density in excitations per $\mu$m vs.\ position in the 1d trap and detuning. Dashed line: resonance line corresponding to $2\Delta=C_6/r^6$ as explained in the text. (b) Excitation density for $\Delta/\Omega=7$ split into excitation-number subspaces.}
 \label{fig:Nex_z}
\end{figure}

\subsubsection{Position-resolved Rydberg excitation density}

The sharply peaked $g^{(2)}$-function in Fig.~\ref{fig:g2} indicated that the distances between excitations predominantly are multiples of the resonant  excitation distances. But interestingly, this does not translate into a peaked structure in the spatially resolved excitation density. Fig.~\ref{fig:Nex_z}(a) shows that at low detuning ($\Delta\approx\Omega$), as is well-known \cite{gaerttner2012}, an enhancement of the 
population occurs close to the trap edges. In a simple blockade picture, this is due to the fact that for an atom close to the trap edge, the probability to be blockaded by a nearby excitation is smaller since less atoms are present in its vicinity. At high detuning, the excitation density has a rather complicated step-like dependence on the position and the trap edges become depleted. This can be explained with a geometric argument: If we consider, e.g., the doubly excited states, we notice that these are only populated if the distance between the excited atoms is the resonant one. If we now ask how such a pair of excitations with fixed distance can be placed in the 1d ensemble volume, and assume that each of these possibilities is realized with equal probability, it becomes clear that atoms located within one resonant radius from the  ends of the trap are excited only half as often as atoms at the center. This explains the outermost edges in the excitation density. The dashed line in Fig.~\ref{fig:Nex_z}(a) indicates the position  $r\ind{res,2}$ away from one edge of the ensemble coinciding with the position of the step in excitation density, clearly supporting this interpretation. The other edges can be explained analogously, taking into account higher excited states. This is also illustrated in Fig.~\ref{fig:Nex_z}(b) in which the excitation density is split up into contributions of states with different excitation numbers. Again, e.g., the plateau-structure of the doubly excited states confirms the above reasoning. In higher dimensions the condition of resonant excitation is fulfilled for various positions of the third excitation and thus it is less localized.
These results indicate a second fundamental difference of the regular structures found in our setup from previous proposals. The GSC are located at fixed positions relative to the trap borders. In our case, however, the resonant excitation structures are not fixed relative to the ensemble, but can \emph{float} over a certain position range in the atom gas from realization to realization. This leads to the characteristic differences in the excitation density shown in Fig.~\ref{fig:Nex_z}.

\subsection{Dynamical buildup of correlations}
\label{sec:g2tdep}

\begin{figure}[t]
 \centering
 \includegraphics[width=\columnwidth]{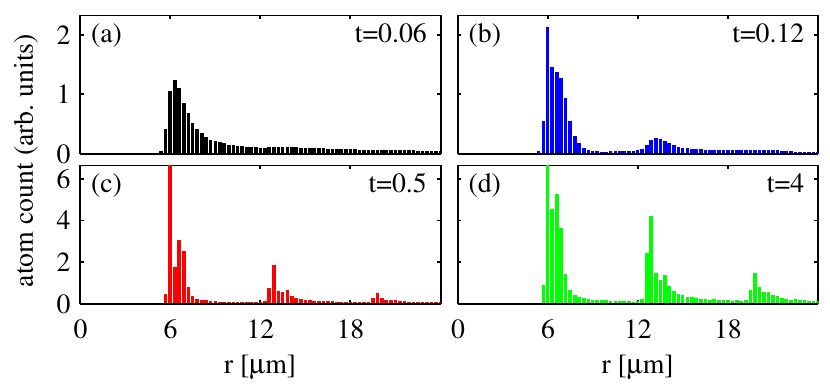}
 \caption{(Color online) Illustration of the aggregate formation: Positions of the excitations are determined from the exact quantum state by a Monte Carlo procedure mimicking a measurement. The distances of the excitations to the leftmost excitation are collected in a histogram. Note that the scale of the ordinate in (c) and (d) differs from (a) and (b). The height of the two-photon resonance is 12 in (c) and 19 in (d). Parameters are as in Fig.~\ref{fig:g2}, $\Delta/\Omega=6$. In (d) the steady state has been reached. Excitation times are given in $\mu$s.}
 \label{fig:res_width_cryst}
\end{figure}

To reveal the emergence of spatial order in our setup, a measurement as illustrated in Fig.~\ref{fig:res_width_cryst} could be performed. In each run, a certain excitation pattern is obtained. To compensate for the floating, the positions of the leftmost excited atoms in the different runs are matched by a shift of the position axis. Then, the shifted data of the different runs are averaged. A histogram of 2000  runs analyzed in this way is shown in Fig.~\ref{fig:res_width_cryst} and clearly shows the strongly peaked structure at long times. At short excitation times the correlation peaks build up successively. Remarkably, after 500\,ns the qualitative features of the steady-state situation are present already. Thereafter, only the absolute height of the peaks increases. This successive buildup of correlation peaks can be attributed to a finite propagation speed of entanglement, that was studied in lattice geometries recently \cite{schachenmayer2013}.

It can be seen from Fig.~\ref{fig:res_width_cryst} that the correlations decrease with increasing distance, unlike in an ideal crystal. But the correlations extend over a multiple of the interaction length scale. In consistence with our interpretations, we found that the characteristic distances between excitations are independent of the ensemble volume.

\section{Quantum gate with asymmetric structures}
\label{sec:gate}

We now turn to a specific application exploiting the spatially asymmetric excitation structure, and show that the generated spatial structures are optimal in the sense that they maximize the success probability of our proposed application. For this we consider an implementation of a three-particle quantum gate~\cite{saffman2010,saffman2005,protsenko2002,isenhower2010}. Based on the approach used in Ref.~\cite{schauss2012}, the excitation structure generated in the first step can be isolated by removing ground state atoms with a resonant laser pulse, and subsequently mapped onto ground states by resonantly driving a transition to a rapidly decaying $p$-state. That way, a spatial arrangement of ground state atoms is prepared, that corresponds to one specific realization of the original excitation structure. In the case that three atoms survive this procedure, their mutual distances are $r\ind{res,2}$, $r\ind{res,1}$, and $r\ind{res,2}+r\ind{res,1}$, and we denote the three atoms as $C$, $A$ and $B$ as 
indicated in Fig.~\ref{fig:Nex_z}(a), respectively.

\begin{figure}[t]
 \centering
 \includegraphics[width=\columnwidth]{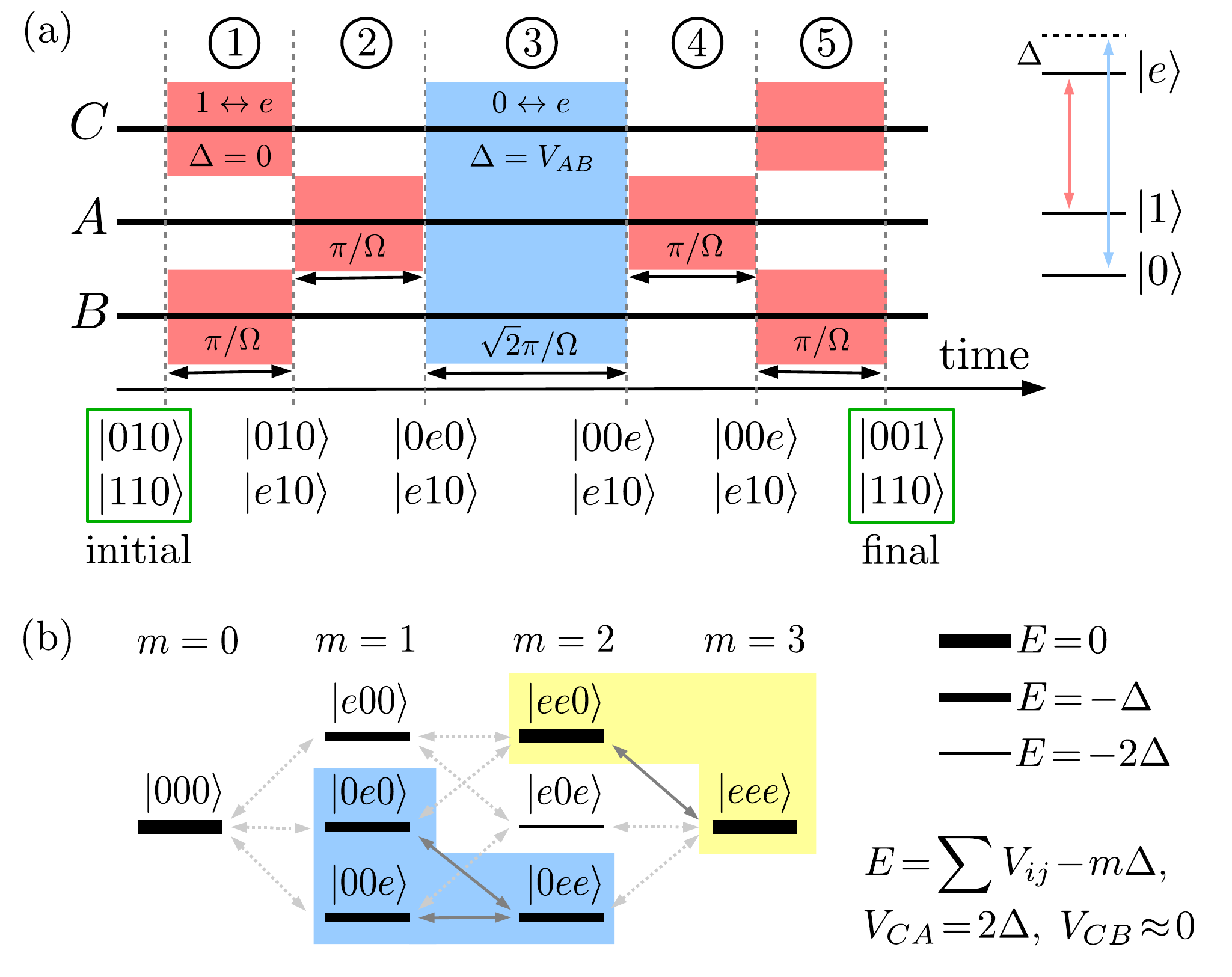}
 \caption{(Color online) (a) Illustration of the gate protocol and level scheme of a single atom. The truth table is given for two exemplary initial states $\ket{CAB}=\ket{010}$ and $\ket{110}$. Depending on the control bit $C$ bits $A$ and $B$ interchange their states, or not. (b) Hamiltonian of the detuned pulse (blue in gate protocol). The thickness of the bars stands for the value $E$ of the diagonal element of the corresponding state (with $m$ Rydberg excitations). Double arrows indicate laser couplings. In the blue subspace, the state $\ket{0e0}$ evolves into $\ket{00e}$ under the $\sqrt{2}\pi$-pulse.}
 \label{fig:gate}
\end{figure}

We exploit this asymmetric arrangement of atoms to implement a controlled SWAP (CSWAP) or Fredkin gate. Depending on the state of the control atom $C$, atoms $A$ and $B$ shall interchange their states or not. The qubits are stored in two hyperfine components of the ground state $\ket{0}$ and $\ket{1}$ that can both be coupled to the Rydberg state $\ket{e}$. The gate protocol consisting of five laser pulses is illustrated in Fig.~\ref{fig:gate}(a). As an example, two possible evolution sequences for initial states $\ket{CAB} = \ket{010}$ or $\ket{110}$ out of the full truth table are shown in the bottom part of Fig.~\ref{fig:gate}(a). 

In step 1, a resonant $\pi$-pulse on atoms $C$ and $B$ evolves them to $|e\rangle$ if they are initially in $\ket{1}$, but leaves them untouched if they are in $\ket{0}$. 
Step 2 is a resonant $\pi$-pulse on atom $A$. Again, it induces excitations of $A$ from $\ket{1}$, but not from $\ket{0}$. But due to the excitation blockade, in addition the excitation only occurs if neither $C$ nor $B$ are excited. Therefore, atom $A$ is excited in this step for initial state $\ket{010}$, but not for initial state $\ket{110}$, see Fig.~\ref{fig:gate}(a).
The crucial step is the detuned $\sqrt{2}\pi$-pulse (step 3) in the middle of the sequence applied to all three atoms. The corresponding effective Hamiltonian is shown schematically in Fig.~\ref{fig:gate}(b). Because of the detuning, it decomposes into several approximately independent subspaces. Due to this separation, its main effect is an interchange of states $\ket{0e0}$ and $\ket{00e}$. Besides this desired exchange, also $\ket{ee0}$ would be resonantly coupled to $\ket{eee}$ causing unwanted dynamics. But due to the blockade in the second step,  $\ket{ee0}$ and $\ket{eee}$ are never accessed, such that this channel can be neglected.
Steps 4 and 5 repeat the first two steps in reverse order, and effectively evolve the atoms back into a superposition of ground states $\ket{0}$ and $\ket{1}$.

\begin{figure}[t]
 \centering
 \includegraphics[width=\columnwidth]{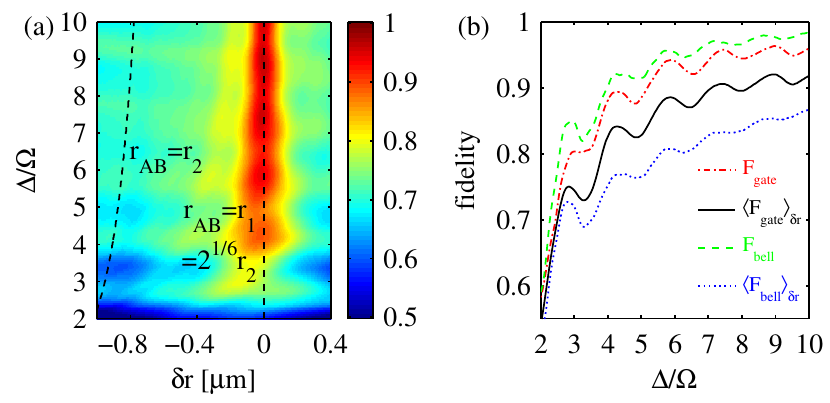}
 \caption{(Color online) (a) Fidelity $F\ind{gate}$ of the CSWAP gate against detuning $\Delta$ and imperfections $\delta r$ of the spatial arrangement of the three atoms. (b) $F\ind{gate}$ and $F\ind{bell}$ for the optimal $r_{AB}$ and averaged over a Lorentzian distribution of distances $r_{AB}$. The fidelity is relatively insensitive to variations in $r_{CA}$.}
 \label{fig:qinfo}
\end{figure}

We have implemented this five-step sequence on the full three-atom state space and numerically simulated the dynamics.
To quantify the quality of the gate for arbitrary input states, we use the gate fidelity 
\begin{align}
F\ind{gate}= \frac 18 \sum_i|\langle\Psi^{(i)}|\Psi\ind{id}^{(i)}\rangle|^2\,,
\end{align}
 where $\Psi\ind{id}^{(i)}$ and $\Psi^{(i)}$ are the ideal target and the numerically obtained output state, respectively, as well as the confidence $F\ind{bell}$ with which a maximally entangled state $\ket{\Psi}\ind{bell}=(\ket{001}-\ket{110})/\sqrt{2}$ can be prepared from the unentangled initial state $(\ket{010}+\ket{110})/\sqrt{2}$. 
Figure~\ref{fig:qinfo}(a) shows $F\ind{gate}$ as a function of $\Delta/\Omega$ and the deviation $\delta r$ of $r_{AB}$ from $r\ind{res,1}$. The fidelity increases with $\Delta/\Omega$ since the decomposition of the Hamiltonian into independent subspaces improves with $\Delta$. We also find that $F\ind{gate}$ is very sensitive to variations in $r_{AB}$, and reaches the optimum value only for $\delta r = 0$. 
In Fig.~\ref{fig:qinfo}(b) results for both, $F\ind{gate}$ and $F\ind{bell}$, with optimized $r_{AB}$ and integrated over the distribution of distances $r_{AB}$ that results from the preparation using the resonant excitation scheme are shown. We notice that the entanglement fidelity is even more sensitive to variations in $r_{AB}$ than the gate fidelity. The reason is that the Bell state preparation strongly depends on the relative phases that the different product states acquire during the gate operation. These phases are more difficult to control, the higher $\Delta/\Omega$ becomes. Thus they counteract the decreasing coupling between the subspaces of the Hamiltonian and decrease the fidelity.

It is crucial to realize that the required separation of the Hamiltonian into different subspaces only arises due to the asymmetric distances $r_{CA}/r_{AB}\neq 1$. Furthermore, already small deviations in the optimum distance $\delta r$ lead to a significant degradation of the gate fidelity. In our setup this optimum distance is automatically achieved for each given laser detuning, because the resonantly excited correlated structures are created with atomic distances such that the laser detuning condition required for the gate operation is satisfied. It is in this sense, that the self-assembly of the excitation structures is optimal.

An obvious drawback of the proposed scheme is, that after generating the resonantly excited triples out of a homogeneous sample, one does not know where exactly the atoms are localized due to the floating nature of the excitation structures. This will spoil the required single atom addressability. A solution could be to use a dense sample in a small trap that accommodates at most 3 excitations, such that the range for the floating is smaller than the distances between the excitations. However, this still leaves two possible orientations for the created resonant structure. Alternatively, one could use a lattice geometry for the whole protocol. This would have the advantage that after preparing the asymmetric structure, one could determine the positions of the individual atoms, e.g., by fluorescence imaging. The disadvantage would be that the irrational distance ratio of $2^{1/6}$ is difficult to realize.
If single site addressability is achieved, the initial state preparation can be realized by coupling the two hyperfine states by a microwave field. For readout, one could employ state selective fluorescence imaging, as described in Ref.~\cite{schauss2012}.

The preparation of the atom structure and the gate protocol itself could further be affected by atomic motion, e.g., induced by mechanical repulsion between the constituents of a resonantly excited pair. It has been show recently that this can deteriorate the resonant pair excitation mechanism \cite{li2013}. In our case it would probably lead to a further broadening of the position space resonances and also to additional dephasing effects~\cite{gaerttner2013} during the gate operation, since for the initial state $\ket{001}$ and $\ket{010}$, the state $\ket{0ee}$ is temporarily populated, which suffers from the repulsion between atoms $A$ and $B$. The consistent inclusion of motional degrees of freedom in a many-body system of long range interacting atoms, however, remains an unresolved challenge.

\section{Discussion and summary}

In summary, we have shown that regular Rydberg excitation structures form at off-resonant laser driving. These structures differ from previously studied ground state crystals, as their characteristic distances are independent of the ensemble length, as they are not localized relative to the gas, and as the Mandel $Q$ parameter, the Rydberg excitation density and the total number of Rydberg excitations yield qualitatively different results. We have shown how these structures build up dynamically and found that the ratio of the two different emerging characteristic lengths depends on the interaction potential only.

As an application, we have shown that excitation structures generated by off-resonant laser driving can be used to implement an efficient three-qubit quantum gate, because the excitation scheme automatically generates an asymmetric spatial arrangement out of a homogeneous atom cloud such that an optimum gate operation is achieved for the given laser parameters. This asymmetry that the gate protocol relies on, is a feature that is inherent to resonantly generated structures and cannot be found in 
ground state crystals. Extensions to other gates like CNOT or Toffoli are straight forward. We also expect that the general concept of self-assembly of optimum asymmetric excitation structures will find other applications.

Experimentally, laser excited Rydberg atoms in a quasi one-dimensional dipole trap could be used to verify our theoretical predictions. Depending on the capabilities of the experimental setup, different predictions can be probed. As the simplest observable, the total number of excited atoms could be measured as a function of the detuning and the density. This way, the shift of the maximum of the Rydberg excitation to higher detunings with increasing density and the increasingly super-Poissonian excitation statistics at very high detuning could be verified as a first manifestation of the resonant excitation channels. 
A position-resolved measurement of the excitations would allow to verify the predicted resonance peaks in $g^{(2)}$, and the formation of the ordered structure as described in Sec.~\ref{sec:g2tdep}. By determining the position of the resonances as a function of the detuning, also the precise structure of the interaction potential could be probed. 
%
Furthermore, the spatial dependence of the various excitation-number subspaces arising from the floating nature of our aggregates could be analyzed. Realizing an excitation volume with sharp boundaries, for example by using flat top beams, one could investigate the enhanced excitation density close to the trap border  in the case of resonant excitation which is suppressed in the far off-resonant case (see Fig.~\ref{fig:Nex_z}).
The resonant excitation aggregates are formed dynamically throughout the time evolution and do not require a thermalization of the system. This invites a time-resolved study of the dynamic formation of spatial correlations.

In approximately 1D systems or higher-dimensional systems, the higher-order resonance conditions do not uniquely determine the positions of all involved particles. This leads to a broadening of the corresponding resonances in $g^{(2)}$. 
Also decoherence due to spontaneous emission, finite laser linewidth, or effects of atomic motion will lead to a broadening of the resonance peaks in $g^{(2)}$. All these effects lead to a suppression of the resonant two photon processes compared to the off-resonant single photon excitation, which sets limits to the observation of higher-order resonance peaks in experiments.

Finally, we note that  very recently, super-Poissonian excitation statistics have been observed in experiments operating at off-resonant driving in the strongly dissipative regime \cite{schempp2013,malossi2013}. We expect that the mechanism of aggregate formation discussed here leading to a bimodal distribution of excitation numbers still applies in the presence of substantial decoherence. The main difference would be that, rather than resonantly excited pairs, off-resonantly excited single atoms function as initial grains for the aggregate formation~\cite{schempp2013}.

\begin{acknowledgments}
We gratefully acknowledge helpful discussions with 
G.\ G\"unter, 
C.\ S.\ Hofmann,
O.\ Morsch,
M.\ Robert-de-Saint-Vincent, 
H.\ Schempp, 
D.\ W.\ Sch\"onleber,
M.\ Weidem\"uller, and 
S.\ Whitlock.
This work was supported by University of Heidelberg (Center for Quantum Dynamics, LGFG), by Deutsche Forschungsgemeinschaft (GA 677/7,8), and  
by the Helmholtz Association (HA216/EMMI).
\end{acknowledgments}


\begin{thebibliography}{41}%
\makeatletter
\providecommand \@ifxundefined [1]{%
 \@ifx{#1\undefined}
}%
\providecommand \@ifnum [1]{%
 \ifnum #1\expandafter \@firstoftwo
 \else \expandafter \@secondoftwo
 \fi
}%
\providecommand \@ifx [1]{%
 \ifx #1\expandafter \@firstoftwo
 \else \expandafter \@secondoftwo
 \fi
}%
\providecommand \natexlab [1]{#1}%
\providecommand \enquote  [1]{``#1''}%
\providecommand \bibnamefont  [1]{#1}%
\providecommand \bibfnamefont [1]{#1}%
\providecommand \citenamefont [1]{#1}%
\providecommand \href@noop [0]{\@secondoftwo}%
\providecommand \href [0]{\begingroup \@sanitize@url \@href}%
\providecommand \@href[1]{\@@startlink{#1}\@@href}%
\providecommand \@@href[1]{\endgroup#1\@@endlink}%
\providecommand \@sanitize@url [0]{\catcode `\\12\catcode `\$12\catcode
  `\&12\catcode `\#12\catcode `\^12\catcode `\_12\catcode `\%12\relax}%
\providecommand \@@startlink[1]{}%
\providecommand \@@endlink[0]{}%
\providecommand \url  [0]{\begingroup\@sanitize@url \@url }%
\providecommand \@url [1]{\endgroup\@href {#1}{\urlprefix }}%
\providecommand \urlprefix  [0]{URL }%
\providecommand \Eprint [0]{\href }%
\providecommand \doibase [0]{http://dx.doi.org/}%
\providecommand \selectlanguage [0]{\@gobble}%
\providecommand \bibinfo  [0]{\@secondoftwo}%
\providecommand \bibfield  [0]{\@secondoftwo}%
\providecommand \translation [1]{[#1]}%
\providecommand \BibitemOpen [0]{}%
\providecommand \bibitemStop [0]{}%
\providecommand \bibitemNoStop [0]{.\EOS\space}%
\providecommand \EOS [0]{\spacefactor3000\relax}%
\providecommand \BibitemShut  [1]{\csname bibitem#1\endcsname}%
\let\auto@bib@innerbib\@empty
\bibitem [{\citenamefont {Saffman}\ \emph {et~al.}(2010)\citenamefont
  {Saffman}, \citenamefont {Walker},\ and\ \citenamefont
  {M\o{}lmer}}]{saffman2010}%
  \BibitemOpen
  \bibfield  {author} {\bibinfo {author} {\bibfnamefont {M.}~\bibnamefont
  {Saffman}}, \bibinfo {author} {\bibfnamefont {T.~G.}\ \bibnamefont {Walker}},
  \ and\ \bibinfo {author} {\bibfnamefont {K.}~\bibnamefont {M\o{}lmer}},\
  }\href {\doibase 10.1103/RevModPhys.82.2313} {\bibfield  {journal} {\bibinfo
  {journal} {Rev. Mod. Phys.}\ }\textbf {\bibinfo {volume} {82}},\ \bibinfo
  {pages} {2313} (\bibinfo {year} {2010})}\BibitemShut {NoStop}%
\bibitem [{\citenamefont {Comparat}\ and\ \citenamefont
  {Pillet}(2010)}]{comparat2010}%
  \BibitemOpen
  \bibfield  {author} {\bibinfo {author} {\bibfnamefont {D.}~\bibnamefont
  {Comparat}}\ and\ \bibinfo {author} {\bibfnamefont {P.}~\bibnamefont
  {Pillet}},\ }\href {\doibase 10.1364/JOSAB.27.00A208} {\bibfield  {journal}
  {\bibinfo  {journal} {J. Opt. Soc. Am. B}\ }\textbf {\bibinfo {volume}
  {27}},\ \bibinfo {pages} {A208} (\bibinfo {year} {2010})}\BibitemShut
  {NoStop}%
\bibitem [{\citenamefont {Lukin}\ \emph {et~al.}(2001)\citenamefont {Lukin},
  \citenamefont {Fleischhauer}, \citenamefont {C\^ot\'e}, \citenamefont {Duan},
  \citenamefont {Jaksch}, \citenamefont {Cirac},\ and\ \citenamefont
  {Zoller}}]{lukin2001}%
  \BibitemOpen
  \bibfield  {author} {\bibinfo {author} {\bibfnamefont {M.~D.}\ \bibnamefont
  {Lukin}}, \bibinfo {author} {\bibfnamefont {M.}~\bibnamefont {Fleischhauer}},
  \bibinfo {author} {\bibfnamefont {R.}~\bibnamefont {C\^ot\'e}}, \bibinfo
  {author} {\bibfnamefont {L.~M.}\ \bibnamefont {Duan}}, \bibinfo {author}
  {\bibfnamefont {D.}~\bibnamefont {Jaksch}}, \bibinfo {author} {\bibfnamefont
  {J.~I.}\ \bibnamefont {Cirac}}, \ and\ \bibinfo {author} {\bibfnamefont
  {P.}~\bibnamefont {Zoller}},\ }\href {\doibase 10.1103/PhysRevLett.87.037901}
  {\bibfield  {journal} {\bibinfo  {journal} {Phys. Rev. Lett.}\ }\textbf
  {\bibinfo {volume} {87}},\ \bibinfo {pages} {037901} (\bibinfo {year}
  {2001})}\BibitemShut {NoStop}%
\bibitem [{\citenamefont {Jaksch}\ \emph {et~al.}(2000)\citenamefont {Jaksch},
  \citenamefont {Cirac}, \citenamefont {Zoller}, \citenamefont {Rolston},
  \citenamefont {C\^ot\'e},\ and\ \citenamefont {Lukin}}]{jaksch2000}%
  \BibitemOpen
  \bibfield  {author} {\bibinfo {author} {\bibfnamefont {D.}~\bibnamefont
  {Jaksch}}, \bibinfo {author} {\bibfnamefont {J.~I.}\ \bibnamefont {Cirac}},
  \bibinfo {author} {\bibfnamefont {P.}~\bibnamefont {Zoller}}, \bibinfo
  {author} {\bibfnamefont {S.~L.}\ \bibnamefont {Rolston}}, \bibinfo {author}
  {\bibfnamefont {R.}~\bibnamefont {C\^ot\'e}}, \ and\ \bibinfo {author}
  {\bibfnamefont {M.~D.}\ \bibnamefont {Lukin}},\ }\href {\doibase
  10.1103/PhysRevLett.85.2208} {\bibfield  {journal} {\bibinfo  {journal}
  {Phys. Rev. Lett.}\ }\textbf {\bibinfo {volume} {85}},\ \bibinfo {pages}
  {2208} (\bibinfo {year} {2000})}\BibitemShut {NoStop}%
\bibitem [{\citenamefont {Schwarzkopf}\ \emph {et~al.}(2011)\citenamefont
  {Schwarzkopf}, \citenamefont {Sapiro},\ and\ \citenamefont
  {Raithel}}]{schwarzkopf2011}%
  \BibitemOpen
  \bibfield  {author} {\bibinfo {author} {\bibfnamefont {A.}~\bibnamefont
  {Schwarzkopf}}, \bibinfo {author} {\bibfnamefont {R.~E.}\ \bibnamefont
  {Sapiro}}, \ and\ \bibinfo {author} {\bibfnamefont {G.}~\bibnamefont
  {Raithel}},\ }\href {\doibase 10.1103/PhysRevLett.107.103001} {\bibfield
  {journal} {\bibinfo  {journal} {Phys. Rev. Lett.}\ }\textbf {\bibinfo
  {volume} {107}},\ \bibinfo {pages} {103001} (\bibinfo {year}
  {2011})}\BibitemShut {NoStop}%
\bibitem [{\citenamefont {Schau\ss}\ \emph {et~al.}(2012)\citenamefont
  {Schau\ss}, \citenamefont {Cheneau}, \citenamefont {Endres}, \citenamefont
  {Fukuhara}, \citenamefont {Hild}, \citenamefont {Omran}, \citenamefont
  {Pohl}, \citenamefont {Gross}, \citenamefont {Kuhr},\ and\ \citenamefont
  {Bloch}}]{schauss2012}%
  \BibitemOpen
  \bibfield  {author} {\bibinfo {author} {\bibfnamefont {P.}~\bibnamefont
  {Schau\ss}}, \bibinfo {author} {\bibfnamefont {M.}~\bibnamefont {Cheneau}},
  \bibinfo {author} {\bibfnamefont {M.}~\bibnamefont {Endres}}, \bibinfo
  {author} {\bibfnamefont {T.}~\bibnamefont {Fukuhara}}, \bibinfo {author}
  {\bibfnamefont {S.}~\bibnamefont {Hild}}, \bibinfo {author} {\bibfnamefont
  {A.}~\bibnamefont {Omran}}, \bibinfo {author} {\bibfnamefont
  {T.}~\bibnamefont {Pohl}}, \bibinfo {author} {\bibfnamefont {C.}~\bibnamefont
  {Gross}}, \bibinfo {author} {\bibfnamefont {S.}~\bibnamefont {Kuhr}}, \ and\
  \bibinfo {author} {\bibfnamefont {I.}~\bibnamefont {Bloch}},\ }\href
  {http://dx.doi.org/10.1038/nature11596} {\bibfield  {journal} {\bibinfo
  {journal} {Nature}\ }\textbf {\bibinfo {volume} {491}},\ \bibinfo {pages}
  {87} (\bibinfo {year} {2012})}\BibitemShut {NoStop}%
\bibitem [{\citenamefont {Olmos}\ and\ \citenamefont
  {Lesanovsky}(2011)}]{olmos2011}%
  \BibitemOpen
  \bibfield  {author} {\bibinfo {author} {\bibfnamefont {B.}~\bibnamefont
  {Olmos}}\ and\ \bibinfo {author} {\bibfnamefont {I.}~\bibnamefont
  {Lesanovsky}},\ }\href {\doibase 10.1039/C0CP01451F} {\bibfield  {journal}
  {\bibinfo  {journal} {Phys. Chem. Chem. Phys.}\ }\textbf {\bibinfo {volume}
  {13}},\ \bibinfo {pages} {4208} (\bibinfo {year} {2011})}\BibitemShut
  {NoStop}%
\bibitem [{\citenamefont {G\"{u}nter}\ \emph {et~al.}(2012)\citenamefont
  {G\"{u}nter}, \citenamefont {Robert-de Saint-Vincent}, \citenamefont
  {Schempp}, \citenamefont {Hofmann}, \citenamefont {Whitlock},\ and\
  \citenamefont {Weidem\"{u}ller}}]{guenter2012}%
  \BibitemOpen
  \bibfield  {author} {\bibinfo {author} {\bibfnamefont {G.}~\bibnamefont
  {G\"{u}nter}}, \bibinfo {author} {\bibfnamefont {M.}~\bibnamefont {Robert-de
  Saint-Vincent}}, \bibinfo {author} {\bibfnamefont {H.}~\bibnamefont
  {Schempp}}, \bibinfo {author} {\bibfnamefont {C.~S.}\ \bibnamefont
  {Hofmann}}, \bibinfo {author} {\bibfnamefont {S.}~\bibnamefont {Whitlock}}, \
  and\ \bibinfo {author} {\bibfnamefont {M.}~\bibnamefont {Weidem\"{u}ller}},\
  }\href {\doibase 10.1103/PhysRevLett.108.013002} {\bibfield  {journal}
  {\bibinfo  {journal} {Phys. Rev. Lett.}\ }\textbf {\bibinfo {volume} {108}},\
  \bibinfo {pages} {013002} (\bibinfo {year} {2012})}\BibitemShut {NoStop}%
\bibitem [{\citenamefont {Ates}\ \emph
  {et~al.}(2007{\natexlab{a}})\citenamefont {Ates}, \citenamefont {Pohl},
  \citenamefont {Pattard},\ and\ \citenamefont {Rost}}]{ates2007a}%
  \BibitemOpen
  \bibfield  {author} {\bibinfo {author} {\bibfnamefont {C.}~\bibnamefont
  {Ates}}, \bibinfo {author} {\bibfnamefont {T.}~\bibnamefont {Pohl}}, \bibinfo
  {author} {\bibfnamefont {T.}~\bibnamefont {Pattard}}, \ and\ \bibinfo
  {author} {\bibfnamefont {J.~M.}\ \bibnamefont {Rost}},\ }\href {\doibase
  10.1103/PhysRevA.76.013413} {\bibfield  {journal} {\bibinfo  {journal} {Phys.
  Rev. A}\ }\textbf {\bibinfo {volume} {76}},\ \bibinfo {pages} {013413}
  (\bibinfo {year} {2007}{\natexlab{a}})}\BibitemShut {NoStop}%
\bibitem [{\citenamefont {Heeg}\ \emph {et~al.}(2012)\citenamefont {Heeg},
  \citenamefont {G\"arttner},\ and\ \citenamefont {Evers}}]{heeg2012}%
  \BibitemOpen
  \bibfield  {author} {\bibinfo {author} {\bibfnamefont {K.~P.}\ \bibnamefont
  {Heeg}}, \bibinfo {author} {\bibfnamefont {M.}~\bibnamefont {G\"arttner}}, \
  and\ \bibinfo {author} {\bibfnamefont {J.}~\bibnamefont {Evers}},\ }\href
  {\doibase 10.1103/PhysRevA.86.063421} {\bibfield  {journal} {\bibinfo
  {journal} {Phys. Rev. A}\ }\textbf {\bibinfo {volume} {86}},\ \bibinfo
  {pages} {063421} (\bibinfo {year} {2012})}\BibitemShut {NoStop}%
\bibitem [{\citenamefont {Petrosyan}\ \emph {et~al.}(2013)\citenamefont
  {Petrosyan}, \citenamefont {H\"oning},\ and\ \citenamefont
  {Fleischhauer}}]{petrosyan2012b}%
  \BibitemOpen
  \bibfield  {author} {\bibinfo {author} {\bibfnamefont {D.}~\bibnamefont
  {Petrosyan}}, \bibinfo {author} {\bibfnamefont {M.}~\bibnamefont {H\"oning}},
  \ and\ \bibinfo {author} {\bibfnamefont {M.}~\bibnamefont {Fleischhauer}},\
  }\href {\doibase 10.1103/PhysRevA.87.053414} {\bibfield  {journal} {\bibinfo
  {journal} {Phys. Rev. A}\ }\textbf {\bibinfo {volume} {87}},\ \bibinfo
  {pages} {053414} (\bibinfo {year} {2013})}\BibitemShut {NoStop}%
\bibitem [{\citenamefont {H\"oning}\ \emph {et~al.}(2013)\citenamefont
  {H\"oning}, \citenamefont {Muth}, \citenamefont {Petrosyan},\ and\
  \citenamefont {Fleischhauer}}]{hoenig2013}%
  \BibitemOpen
  \bibfield  {author} {\bibinfo {author} {\bibfnamefont {M.}~\bibnamefont
  {H\"oning}}, \bibinfo {author} {\bibfnamefont {D.}~\bibnamefont {Muth}},
  \bibinfo {author} {\bibfnamefont {D.}~\bibnamefont {Petrosyan}}, \ and\
  \bibinfo {author} {\bibfnamefont {M.}~\bibnamefont {Fleischhauer}},\ }\href
  {\doibase 10.1103/PhysRevA.87.023401} {\bibfield  {journal} {\bibinfo
  {journal} {Phys. Rev. A}\ }\textbf {\bibinfo {volume} {87}},\ \bibinfo
  {pages} {023401} (\bibinfo {year} {2013})}\BibitemShut {NoStop}%
\bibitem [{\citenamefont {Robicheaux}\ and\ \citenamefont
  {Hern\'andez}(2005)}]{robicheaux2005}%
  \BibitemOpen
  \bibfield  {author} {\bibinfo {author} {\bibfnamefont {F.}~\bibnamefont
  {Robicheaux}}\ and\ \bibinfo {author} {\bibfnamefont {J.~V.}\ \bibnamefont
  {Hern\'andez}},\ }\href {\doibase 10.1103/PhysRevA.72.063403} {\bibfield
  {journal} {\bibinfo  {journal} {Phys. Rev. A}\ }\textbf {\bibinfo {volume}
  {72}},\ \bibinfo {pages} {063403} (\bibinfo {year} {2005})}\BibitemShut
  {NoStop}%
\bibitem [{\citenamefont {Ates}\ \emph {et~al.}(2006)\citenamefont {Ates},
  \citenamefont {Pohl}, \citenamefont {Pattard},\ and\ \citenamefont
  {Rost}}]{ates2006}%
  \BibitemOpen
  \bibfield  {author} {\bibinfo {author} {\bibfnamefont {C.}~\bibnamefont
  {Ates}}, \bibinfo {author} {\bibfnamefont {T.}~\bibnamefont {Pohl}}, \bibinfo
  {author} {\bibfnamefont {T.}~\bibnamefont {Pattard}}, \ and\ \bibinfo
  {author} {\bibfnamefont {J.~M.}\ \bibnamefont {Rost}},\ }\href
  {http://iopscience.iop.org/0953-4075/39/11/L02/} {\bibfield  {journal}
  {\bibinfo  {journal} {J. Phys. B}\ }\textbf {\bibinfo {volume} {39}},\
  \bibinfo {pages} {L233 } (\bibinfo {year} {2006})}\BibitemShut {NoStop}%
\bibitem [{\citenamefont {Younge}\ \emph {et~al.}(2009)\citenamefont {Younge},
  \citenamefont {Reinhard}, \citenamefont {Pohl}, \citenamefont {Berman},\ and\
  \citenamefont {Raithel}}]{younge2009}%
  \BibitemOpen
  \bibfield  {author} {\bibinfo {author} {\bibfnamefont {K.~C.}\ \bibnamefont
  {Younge}}, \bibinfo {author} {\bibfnamefont {A.}~\bibnamefont {Reinhard}},
  \bibinfo {author} {\bibfnamefont {T.}~\bibnamefont {Pohl}}, \bibinfo {author}
  {\bibfnamefont {P.~R.}\ \bibnamefont {Berman}}, \ and\ \bibinfo {author}
  {\bibfnamefont {G.}~\bibnamefont {Raithel}},\ }\href {\doibase
  10.1103/PhysRevA.79.043420} {\bibfield  {journal} {\bibinfo  {journal} {Phys.
  Rev. A}\ }\textbf {\bibinfo {volume} {79}},\ \bibinfo {pages} {043420}
  (\bibinfo {year} {2009})}\BibitemShut {NoStop}%
\bibitem [{\citenamefont {Weimer}\ \emph {et~al.}(2008)\citenamefont {Weimer},
  \citenamefont {L\"ow}, \citenamefont {Pfau},\ and\ \citenamefont
  {B\"uchler}}]{weimer2008}%
  \BibitemOpen
  \bibfield  {author} {\bibinfo {author} {\bibfnamefont {H.}~\bibnamefont
  {Weimer}}, \bibinfo {author} {\bibfnamefont {R.}~\bibnamefont {L\"ow}},
  \bibinfo {author} {\bibfnamefont {T.}~\bibnamefont {Pfau}}, \ and\ \bibinfo
  {author} {\bibfnamefont {H.~P.}\ \bibnamefont {B\"uchler}},\ }\href {\doibase
  10.1103/PhysRevLett.101.250601} {\bibfield  {journal} {\bibinfo  {journal}
  {Phys. Rev. Lett.}\ }\textbf {\bibinfo {volume} {101}},\ \bibinfo {pages}
  {250601} (\bibinfo {year} {2008})}\BibitemShut {NoStop}%
\bibitem [{\citenamefont {Weimer}\ and\ \citenamefont
  {B\"uchler}(2010)}]{weimer2010b}%
  \BibitemOpen
  \bibfield  {author} {\bibinfo {author} {\bibfnamefont {H.}~\bibnamefont
  {Weimer}}\ and\ \bibinfo {author} {\bibfnamefont {H.~P.}\ \bibnamefont
  {B\"uchler}},\ }\href {\doibase 10.1103/PhysRevLett.105.230403} {\bibfield
  {journal} {\bibinfo  {journal} {Phys. Rev. Lett.}\ }\textbf {\bibinfo
  {volume} {105}},\ \bibinfo {pages} {230403} (\bibinfo {year}
  {2010})}\BibitemShut {NoStop}%
\bibitem [{\citenamefont {L\"ow}\ \emph {et~al.}(2009)\citenamefont {L\"ow},
  \citenamefont {Weimer}, \citenamefont {Krohn}, \citenamefont {Heidemann},
  \citenamefont {Bendkowsky}, \citenamefont {Butscher}, \citenamefont
  {B\"uchler},\ and\ \citenamefont {Pfau}}]{loew2009}%
  \BibitemOpen
  \bibfield  {author} {\bibinfo {author} {\bibfnamefont {R.}~\bibnamefont
  {L\"ow}}, \bibinfo {author} {\bibfnamefont {H.}~\bibnamefont {Weimer}},
  \bibinfo {author} {\bibfnamefont {U.}~\bibnamefont {Krohn}}, \bibinfo
  {author} {\bibfnamefont {R.}~\bibnamefont {Heidemann}}, \bibinfo {author}
  {\bibfnamefont {V.}~\bibnamefont {Bendkowsky}}, \bibinfo {author}
  {\bibfnamefont {B.}~\bibnamefont {Butscher}}, \bibinfo {author}
  {\bibfnamefont {H.~P.}\ \bibnamefont {B\"uchler}}, \ and\ \bibinfo {author}
  {\bibfnamefont {T.}~\bibnamefont {Pfau}},\ }\href {\doibase
  10.1103/PhysRevA.80.033422} {\bibfield  {journal} {\bibinfo  {journal} {Phys.
  Rev. A}\ }\textbf {\bibinfo {volume} {80}},\ \bibinfo {pages} {033422}
  (\bibinfo {year} {2009})}\BibitemShut {NoStop}%
\bibitem [{\citenamefont {Olmos}\ \emph {et~al.}(2009)\citenamefont {Olmos},
  \citenamefont {Gonz\'alez-F\'erez},\ and\ \citenamefont
  {Lesanovsky}}]{olmos2009a}%
  \BibitemOpen
  \bibfield  {author} {\bibinfo {author} {\bibfnamefont {B.}~\bibnamefont
  {Olmos}}, \bibinfo {author} {\bibfnamefont {R.}~\bibnamefont
  {Gonz\'alez-F\'erez}}, \ and\ \bibinfo {author} {\bibfnamefont
  {I.}~\bibnamefont {Lesanovsky}},\ }\href {\doibase
  10.1103/PhysRevA.79.043419} {\bibfield  {journal} {\bibinfo  {journal} {Phys.
  Rev. A}\ }\textbf {\bibinfo {volume} {79}},\ \bibinfo {pages} {043419}
  (\bibinfo {year} {2009})}\BibitemShut {NoStop}%
\bibitem [{\citenamefont {Tezak}\ \emph {et~al.}(2011)\citenamefont {Tezak},
  \citenamefont {Mayle},\ and\ \citenamefont {Schmelcher}}]{tezak2011}%
  \BibitemOpen
  \bibfield  {author} {\bibinfo {author} {\bibfnamefont {N.}~\bibnamefont
  {Tezak}}, \bibinfo {author} {\bibfnamefont {M.}~\bibnamefont {Mayle}}, \ and\
  \bibinfo {author} {\bibfnamefont {P.}~\bibnamefont {Schmelcher}},\ }\href
  {http://stacks.iop.org/0953-4075/44/i=18/a=184009} {\bibfield  {journal}
  {\bibinfo  {journal} {J. Phys. B}\ }\textbf {\bibinfo {volume} {44}},\
  \bibinfo {pages} {184009} (\bibinfo {year} {2011})}\BibitemShut {NoStop}%
\bibitem [{\citenamefont {Mayle}\ \emph {et~al.}(2011)\citenamefont {Mayle},
  \citenamefont {Zeller}, \citenamefont {Tezak},\ and\ \citenamefont
  {Schmelcher}}]{mayle2011}%
  \BibitemOpen
  \bibfield  {author} {\bibinfo {author} {\bibfnamefont {M.}~\bibnamefont
  {Mayle}}, \bibinfo {author} {\bibfnamefont {W.}~\bibnamefont {Zeller}},
  \bibinfo {author} {\bibfnamefont {N.}~\bibnamefont {Tezak}}, \ and\ \bibinfo
  {author} {\bibfnamefont {P.}~\bibnamefont {Schmelcher}},\ }\href {\doibase
  10.1103/PhysRevA.84.010701} {\bibfield  {journal} {\bibinfo  {journal} {Phys.
  Rev. A}\ }\textbf {\bibinfo {volume} {84}},\ \bibinfo {pages} {010701}
  (\bibinfo {year} {2011})}\BibitemShut {NoStop}%
\bibitem [{\citenamefont {Breyel}\ \emph {et~al.}(2012)\citenamefont {Breyel},
  \citenamefont {Schmidt},\ and\ \citenamefont {Komnik}}]{breyel2012}%
  \BibitemOpen
  \bibfield  {author} {\bibinfo {author} {\bibfnamefont {D.}~\bibnamefont
  {Breyel}}, \bibinfo {author} {\bibfnamefont {T.~L.}\ \bibnamefont {Schmidt}},
  \ and\ \bibinfo {author} {\bibfnamefont {A.}~\bibnamefont {Komnik}},\ }\href
  {\doibase 10.1103/PhysRevA.86.023405} {\bibfield  {journal} {\bibinfo
  {journal} {Phys. Rev. A}\ }\textbf {\bibinfo {volume} {86}},\ \bibinfo
  {pages} {023405} (\bibinfo {year} {2012})}\BibitemShut {NoStop}%
\bibitem [{\citenamefont {Lee}\ \emph {et~al.}(2011)\citenamefont {Lee},
  \citenamefont {H\"affner},\ and\ \citenamefont {Cross}}]{lee2011}%
  \BibitemOpen
  \bibfield  {author} {\bibinfo {author} {\bibfnamefont {T.~E.}\ \bibnamefont
  {Lee}}, \bibinfo {author} {\bibfnamefont {H.}~\bibnamefont {H\"affner}}, \
  and\ \bibinfo {author} {\bibfnamefont {M.~C.}\ \bibnamefont {Cross}},\ }\href
  {\doibase 10.1103/PhysRevA.84.031402} {\bibfield  {journal} {\bibinfo
  {journal} {Phys. Rev. A}\ }\textbf {\bibinfo {volume} {84}},\ \bibinfo
  {pages} {031402} (\bibinfo {year} {2011})}\BibitemShut {NoStop}%
\bibitem [{\citenamefont {Ates}\ \emph
  {et~al.}(2007{\natexlab{b}})\citenamefont {Ates}, \citenamefont {Pohl},
  \citenamefont {Pattard},\ and\ \citenamefont {Rost}}]{ates2007b}%
  \BibitemOpen
  \bibfield  {author} {\bibinfo {author} {\bibfnamefont {C.}~\bibnamefont
  {Ates}}, \bibinfo {author} {\bibfnamefont {T.}~\bibnamefont {Pohl}}, \bibinfo
  {author} {\bibfnamefont {T.}~\bibnamefont {Pattard}}, \ and\ \bibinfo
  {author} {\bibfnamefont {J.~M.}\ \bibnamefont {Rost}},\ }\href {\doibase
  10.1103/PhysRevLett.98.023002} {\bibfield  {journal} {\bibinfo  {journal}
  {Phys. Rev. Lett.}\ }\textbf {\bibinfo {volume} {98}},\ \bibinfo {pages}
  {023002} (\bibinfo {year} {2007}{\natexlab{b}})}\BibitemShut {NoStop}%
\bibitem [{\citenamefont {Amthor}\ \emph {et~al.}(2010)\citenamefont {Amthor},
  \citenamefont {Giese}, \citenamefont {Hofmann},\ and\ \citenamefont
  {Weidem\"uller}}]{amthor2010}%
  \BibitemOpen
  \bibfield  {author} {\bibinfo {author} {\bibfnamefont {T.}~\bibnamefont
  {Amthor}}, \bibinfo {author} {\bibfnamefont {C.}~\bibnamefont {Giese}},
  \bibinfo {author} {\bibfnamefont {C.~S.}\ \bibnamefont {Hofmann}}, \ and\
  \bibinfo {author} {\bibfnamefont {M.}~\bibnamefont {Weidem\"uller}},\ }\href
  {\doibase 10.1103/PhysRevLett.104.013001} {\bibfield  {journal} {\bibinfo
  {journal} {Phys. Rev. Lett.}\ }\textbf {\bibinfo {volume} {104}},\ \bibinfo
  {pages} {013001} (\bibinfo {year} {2010})}\BibitemShut {NoStop}%
\bibitem [{\citenamefont {Pohl}\ \emph {et~al.}(2010)\citenamefont {Pohl},
  \citenamefont {Demler},\ and\ \citenamefont {Lukin}}]{pohl2010}%
  \BibitemOpen
  \bibfield  {author} {\bibinfo {author} {\bibfnamefont {T.}~\bibnamefont
  {Pohl}}, \bibinfo {author} {\bibfnamefont {E.}~\bibnamefont {Demler}}, \ and\
  \bibinfo {author} {\bibfnamefont {M.~D.}\ \bibnamefont {Lukin}},\ }\href
  {\doibase 10.1103/PhysRevLett.104.043002} {\bibfield  {journal} {\bibinfo
  {journal} {Phys. Rev. Lett.}\ }\textbf {\bibinfo {volume} {104}},\ \bibinfo
  {pages} {043002} (\bibinfo {year} {2010})}\BibitemShut {NoStop}%
\bibitem [{\citenamefont {van Bijnen}\ \emph {et~al.}(2011)\citenamefont {van
  Bijnen}, \citenamefont {Smit}, \citenamefont {van Leeuwen}, \citenamefont
  {Vredenbregt},\ and\ \citenamefont {Kokkelmans}}]{vanbijnen2011}%
  \BibitemOpen
  \bibfield  {author} {\bibinfo {author} {\bibfnamefont {R.~M.~W.}\
  \bibnamefont {van Bijnen}}, \bibinfo {author} {\bibfnamefont
  {S.}~\bibnamefont {Smit}}, \bibinfo {author} {\bibfnamefont {K.~A.~H.}\
  \bibnamefont {van Leeuwen}}, \bibinfo {author} {\bibfnamefont {E.~J.~D.}\
  \bibnamefont {Vredenbregt}}, \ and\ \bibinfo {author} {\bibfnamefont {S.~J.
  J. M.~F.}\ \bibnamefont {Kokkelmans}},\ }\href
  {http://iopscience.iop.org/0953-4075/44/18/184008/} {\bibfield  {journal}
  {\bibinfo  {journal} {J. Phys. B}\ }\textbf {\bibinfo {volume} {44}},\
  \bibinfo {pages} {184008} (\bibinfo {year} {2011})}\BibitemShut {NoStop}%
\bibitem [{\citenamefont {Sela}\ \emph {et~al.}(2011)\citenamefont {Sela},
  \citenamefont {Punk},\ and\ \citenamefont {Garst}}]{sela2011}%
  \BibitemOpen
  \bibfield  {author} {\bibinfo {author} {\bibfnamefont {E.}~\bibnamefont
  {Sela}}, \bibinfo {author} {\bibfnamefont {M.}~\bibnamefont {Punk}}, \ and\
  \bibinfo {author} {\bibfnamefont {M.}~\bibnamefont {Garst}},\ }\href
  {\doibase 10.1103/PhysRevB.84.085434} {\bibfield  {journal} {\bibinfo
  {journal} {Phys. Rev. B}\ }\textbf {\bibinfo {volume} {84}},\ \bibinfo
  {pages} {085434} (\bibinfo {year} {2011})}\BibitemShut {NoStop}%
\bibitem [{\citenamefont {Schachenmayer}\ \emph {et~al.}(2010)\citenamefont
  {Schachenmayer}, \citenamefont {Lesanovsky}, \citenamefont {Micheli},\ and\
  \citenamefont {Daley}}]{schachenmayer2010}%
  \BibitemOpen
  \bibfield  {author} {\bibinfo {author} {\bibfnamefont {J.}~\bibnamefont
  {Schachenmayer}}, \bibinfo {author} {\bibfnamefont {I.}~\bibnamefont
  {Lesanovsky}}, \bibinfo {author} {\bibfnamefont {A.}~\bibnamefont {Micheli}},
  \ and\ \bibinfo {author} {\bibfnamefont {A.~J.}\ \bibnamefont {Daley}},\
  }\href {http://stacks.iop.org/1367-2630/12/i=10/a=103044} {\bibfield
  {journal} {\bibinfo  {journal} {New Journal of Physics}\ }\textbf {\bibinfo
  {volume} {491}},\ \bibinfo {pages} {103044} (\bibinfo {year}
  {2010})}\BibitemShut {NoStop}%
\bibitem [{\citenamefont {Lemeshko}\ \emph {et~al.}(2012)\citenamefont
  {Lemeshko}, \citenamefont {Krems},\ and\ \citenamefont
  {Weimer}}]{lemeshko2012}%
  \BibitemOpen
  \bibfield  {author} {\bibinfo {author} {\bibfnamefont {M.}~\bibnamefont
  {Lemeshko}}, \bibinfo {author} {\bibfnamefont {R.~V.}\ \bibnamefont {Krems}},
  \ and\ \bibinfo {author} {\bibfnamefont {H.}~\bibnamefont {Weimer}},\ }\href
  {\doibase 10.1103/PhysRevLett.109.035301} {\bibfield  {journal} {\bibinfo
  {journal} {Phys. Rev. Lett.}\ }\textbf {\bibinfo {volume} {109}},\ \bibinfo
  {pages} {035301} (\bibinfo {year} {2012})}\BibitemShut {NoStop}%
\bibitem [{\citenamefont {G\"arttner}\ \emph {et~al.}(2012)\citenamefont
  {G\"arttner}, \citenamefont {Heeg}, \citenamefont {Gasenzer},\ and\
  \citenamefont {Evers}}]{gaerttner2012}%
  \BibitemOpen
  \bibfield  {author} {\bibinfo {author} {\bibfnamefont {M.}~\bibnamefont
  {G\"arttner}}, \bibinfo {author} {\bibfnamefont {K.~P.}\ \bibnamefont
  {Heeg}}, \bibinfo {author} {\bibfnamefont {T.}~\bibnamefont {Gasenzer}}, \
  and\ \bibinfo {author} {\bibfnamefont {J.}~\bibnamefont {Evers}},\ }\href
  {\doibase 10.1103/PhysRevA.86.033422} {\bibfield  {journal} {\bibinfo
  {journal} {Phys. Rev. A}\ }\textbf {\bibinfo {volume} {86}},\ \bibinfo
  {pages} {033422} (\bibinfo {year} {2012})}\BibitemShut {NoStop}%
\bibitem [{\citenamefont {Carroll}\ \emph {et~al.}(2009)\citenamefont
  {Carroll}, \citenamefont {Daniel}, \citenamefont {Hoover}, \citenamefont
  {Sidie},\ and\ \citenamefont {Noel}}]{carroll2009}%
  \BibitemOpen
  \bibfield  {author} {\bibinfo {author} {\bibfnamefont {T.~J.}\ \bibnamefont
  {Carroll}}, \bibinfo {author} {\bibfnamefont {C.}~\bibnamefont {Daniel}},
  \bibinfo {author} {\bibfnamefont {L.}~\bibnamefont {Hoover}}, \bibinfo
  {author} {\bibfnamefont {T.}~\bibnamefont {Sidie}}, \ and\ \bibinfo {author}
  {\bibfnamefont {M.~W.}\ \bibnamefont {Noel}},\ }\href {\doibase
  10.1103/PhysRevA.80.052712} {\bibfield  {journal} {\bibinfo  {journal} {Phys.
  Rev. A}\ }\textbf {\bibinfo {volume} {80}},\ \bibinfo {pages} {052712}
  (\bibinfo {year} {2009})}\BibitemShut {NoStop}%
\bibitem [{\citenamefont {Ryabtsev}\ \emph {et~al.}(2010)\citenamefont
  {Ryabtsev}, \citenamefont {Tretyakov}, \citenamefont {Beterov}, \citenamefont
  {Entin},\ and\ \citenamefont {Yakshina}}]{ryabtsev2010}%
  \BibitemOpen
  \bibfield  {author} {\bibinfo {author} {\bibfnamefont {I.~I.}\ \bibnamefont
  {Ryabtsev}}, \bibinfo {author} {\bibfnamefont {D.~B.}\ \bibnamefont
  {Tretyakov}}, \bibinfo {author} {\bibfnamefont {I.~I.}\ \bibnamefont
  {Beterov}}, \bibinfo {author} {\bibfnamefont {V.~M.}\ \bibnamefont {Entin}},
  \ and\ \bibinfo {author} {\bibfnamefont {E.~A.}\ \bibnamefont {Yakshina}},\
  }\href {\doibase 10.1103/PhysRevA.82.053409} {\bibfield  {journal} {\bibinfo
  {journal} {Phys. Rev. A}\ }\textbf {\bibinfo {volume} {82}},\ \bibinfo
  {pages} {053409} (\bibinfo {year} {2010})}\BibitemShut {NoStop}%
\bibitem [{\citenamefont {Schachenmayer}\ \emph {et~al.}()\citenamefont
  {Schachenmayer}, \citenamefont {Lanyon}, \citenamefont {Roos}, ,\ and\
  \citenamefont {Daley}}]{schachenmayer2013}%
  \BibitemOpen
  \bibfield  {author} {\bibinfo {author} {\bibfnamefont {J.}~\bibnamefont
  {Schachenmayer}}, \bibinfo {author} {\bibfnamefont {B.~P.}\ \bibnamefont
  {Lanyon}}, \bibinfo {author} {\bibfnamefont {C.~F.}\ \bibnamefont {Roos}}, ,
  \ and\ \bibinfo {author} {\bibfnamefont {A.~J.}\ \bibnamefont {Daley}},\
  }\href@noop {} {}\bibinfo {howpublished} {arXiv:1305.6880
  [cond-mat.quant-gas]}\BibitemShut {NoStop}%
\bibitem [{\citenamefont {Saffman}\ and\ \citenamefont
  {Walker}(2005)}]{saffman2005}%
  \BibitemOpen
  \bibfield  {author} {\bibinfo {author} {\bibfnamefont {M.}~\bibnamefont
  {Saffman}}\ and\ \bibinfo {author} {\bibfnamefont {T.~G.}\ \bibnamefont
  {Walker}},\ }\href {\doibase 10.1103/PhysRevA.72.022347} {\bibfield
  {journal} {\bibinfo  {journal} {Phys. Rev. A}\ }\textbf {\bibinfo {volume}
  {72}},\ \bibinfo {pages} {022347} (\bibinfo {year} {2005})}\BibitemShut
  {NoStop}%
\bibitem [{\citenamefont {Protsenko}\ \emph {et~al.}(2002)\citenamefont
  {Protsenko}, \citenamefont {Reymond}, \citenamefont {Schlosser},\ and\
  \citenamefont {Grangier}}]{protsenko2002}%
  \BibitemOpen
  \bibfield  {author} {\bibinfo {author} {\bibfnamefont {I.~E.}\ \bibnamefont
  {Protsenko}}, \bibinfo {author} {\bibfnamefont {G.}~\bibnamefont {Reymond}},
  \bibinfo {author} {\bibfnamefont {N.}~\bibnamefont {Schlosser}}, \ and\
  \bibinfo {author} {\bibfnamefont {P.}~\bibnamefont {Grangier}},\ }\href
  {\doibase 10.1103/PhysRevA.65.052301} {\bibfield  {journal} {\bibinfo
  {journal} {Phys. Rev. A}\ }\textbf {\bibinfo {volume} {65}},\ \bibinfo
  {pages} {052301} (\bibinfo {year} {2002})}\BibitemShut {NoStop}%
\bibitem [{\citenamefont {Isenhower}\ \emph {et~al.}(2010)\citenamefont
  {Isenhower}, \citenamefont {Urban}, \citenamefont {Zhang}, \citenamefont
  {Gill}, \citenamefont {Henage}, \citenamefont {Johnson}, \citenamefont
  {Walker},\ and\ \citenamefont {Saffman}}]{isenhower2010}%
  \BibitemOpen
  \bibfield  {author} {\bibinfo {author} {\bibfnamefont {L.}~\bibnamefont
  {Isenhower}}, \bibinfo {author} {\bibfnamefont {E.}~\bibnamefont {Urban}},
  \bibinfo {author} {\bibfnamefont {X.~L.}\ \bibnamefont {Zhang}}, \bibinfo
  {author} {\bibfnamefont {A.~T.}\ \bibnamefont {Gill}}, \bibinfo {author}
  {\bibfnamefont {T.}~\bibnamefont {Henage}}, \bibinfo {author} {\bibfnamefont
  {T.~A.}\ \bibnamefont {Johnson}}, \bibinfo {author} {\bibfnamefont {T.~G.}\
  \bibnamefont {Walker}}, \ and\ \bibinfo {author} {\bibfnamefont
  {M.}~\bibnamefont {Saffman}},\ }\href {\doibase
  10.1103/PhysRevLett.104.010503} {\bibfield  {journal} {\bibinfo  {journal}
  {Phys. Rev. Lett.}\ }\textbf {\bibinfo {volume} {104}},\ \bibinfo {pages}
  {010503} (\bibinfo {year} {2010})}\BibitemShut {NoStop}%
\bibitem [{\citenamefont {Li}\ \emph {et~al.}(2013)\citenamefont {Li},
  \citenamefont {Ates},\ and\ \citenamefont {Lesanovsky}}]{li2013}%
  \BibitemOpen
  \bibfield  {author} {\bibinfo {author} {\bibfnamefont {W.}~\bibnamefont
  {Li}}, \bibinfo {author} {\bibfnamefont {C.}~\bibnamefont {Ates}}, \ and\
  \bibinfo {author} {\bibfnamefont {I.}~\bibnamefont {Lesanovsky}},\ }\href
  {\doibase 10.1103/PhysRevLett.110.213005} {\bibfield  {journal} {\bibinfo
  {journal} {Phys. Rev. Lett.}\ }\textbf {\bibinfo {volume} {110}},\ \bibinfo
  {pages} {213005} (\bibinfo {year} {2013})}\BibitemShut {NoStop}%
\bibitem [{\citenamefont {G\"arttner}\ and\ \citenamefont
  {Evers}()}]{gaerttner2013}%
  \BibitemOpen
  \bibfield  {author} {\bibinfo {author} {\bibfnamefont {M.}~\bibnamefont
  {G\"arttner}}\ and\ \bibinfo {author} {\bibfnamefont {J.}~\bibnamefont
  {Evers}},\ }\href@noop {} {}\bibinfo {howpublished} {arXiv:1305.1458
  [physics.atom-ph]}\BibitemShut {NoStop}%
\bibitem [{\citenamefont {Schempp}\ \emph {et~al.}()\citenamefont {Schempp},
  \citenamefont {G\"unter}, \citenamefont {de~Saint-Vincent}, \citenamefont
  {Hofmann}, \citenamefont {Breyel}, \citenamefont {Komnik}, \citenamefont
  {Sch\"onleber}, \citenamefont {G\"arttner}, \citenamefont {Evers},
  \citenamefont {Whitlock},\ and\ \citenamefont {Weidem\"uller}}]{schempp2013}%
  \BibitemOpen
  \bibfield  {author} {\bibinfo {author} {\bibfnamefont {H.}~\bibnamefont
  {Schempp}}, \bibinfo {author} {\bibfnamefont {G.}~\bibnamefont {G\"unter}},
  \bibinfo {author} {\bibfnamefont {M.~R.}\ \bibnamefont {de~Saint-Vincent}},
  \bibinfo {author} {\bibfnamefont {C.~S.}\ \bibnamefont {Hofmann}}, \bibinfo
  {author} {\bibfnamefont {D.}~\bibnamefont {Breyel}}, \bibinfo {author}
  {\bibfnamefont {A.}~\bibnamefont {Komnik}}, \bibinfo {author} {\bibfnamefont
  {D.~W.}\ \bibnamefont {Sch\"onleber}}, \bibinfo {author} {\bibfnamefont
  {M.}~\bibnamefont {G\"arttner}}, \bibinfo {author} {\bibfnamefont
  {J.}~\bibnamefont {Evers}}, \bibinfo {author} {\bibfnamefont
  {S.}~\bibnamefont {Whitlock}}, \ and\ \bibinfo {author} {\bibfnamefont
  {M.}~\bibnamefont {Weidem\"uller}},\ }\href@noop {} {}\bibinfo {howpublished}
  {arXiv:1308.0264 [physics.atom-ph]}\BibitemShut {NoStop}%
\bibitem [{\citenamefont {Malossi}\ \emph {et~al.}()\citenamefont {Malossi},
  \citenamefont {Valado}, \citenamefont {Scotto}, \citenamefont {Huillery},
  \citenamefont {Pillet}, \citenamefont {Ciampini}, \citenamefont {Arimondo},\
  and\ \citenamefont {Morsch}}]{malossi2013}%
  \BibitemOpen
  \bibfield  {author} {\bibinfo {author} {\bibfnamefont {N.}~\bibnamefont
  {Malossi}}, \bibinfo {author} {\bibfnamefont {M.~M.}\ \bibnamefont {Valado}},
  \bibinfo {author} {\bibfnamefont {S.}~\bibnamefont {Scotto}}, \bibinfo
  {author} {\bibfnamefont {P.}~\bibnamefont {Huillery}}, \bibinfo {author}
  {\bibfnamefont {P.}~\bibnamefont {Pillet}}, \bibinfo {author} {\bibfnamefont
  {D.}~\bibnamefont {Ciampini}}, \bibinfo {author} {\bibfnamefont
  {E.}~\bibnamefont {Arimondo}}, \ and\ \bibinfo {author} {\bibfnamefont
  {O.}~\bibnamefont {Morsch}},\ }\href@noop {} {}\bibinfo {howpublished}
  {arXiv:1308.1854 [cond-mat.quant-gas]}\BibitemShut {NoStop}%
\end{thebibliography}
\end{document}